\documentclass[twocolumn,aps,showpacs,amsmath,amssymb,floatfix,superscriptaddress]{revtex4}
\usepackage{graphicx}
\usepackage{dcolumn}
\usepackage{bm}
\usepackage{hyperref} 
\usepackage{latexsym}
\usepackage{float}

\begin{document}
\title{Consensus Formation in Multi-state Majority and Plurality Models}
\author{P.~Chen}
\email{patrick@bu.edu}
\author{S.~Redner}
\email{redner@bu.edu}
\altaffiliation{Permanent address:
Department of Physics, Boston University, Boston, Massachusetts,
02215 USA} \affiliation{Theoretical Division and Center for
Nonlinear Studies, Los Alamos National Laboratory, Los Alamos, New
Mexico 87545}

\begin{abstract}
  
  We study consensus formation in interacting systems that evolve by
  multi-state majority rule and by plurality rule.  In an update event, a
  group of $G$ agents (with $G$ odd), each endowed with an $s$-state spin
  variable, is specified.  For majority rule, all group members adopt the
  local majority state; for plurality rule the group adopts the local
  plurality state.  This update is repeated until a final consensus state is
  generally reached.  In the mean field limit, the consensus time for an
  $N$-spin system increases as $\ln N$ for both majority and plurality rule,
  with an amplitude that depends on $s$ and $G$.  For finite spatial
  dimensions, domains undergo diffusive coarsening in majority rule when $s$
  or $G$ is small.  For larger $s$ and $G$, opinions spread ballistically
  from the few groups with an initial local majority.  For plurality rule,
  there is always diffusive domain coarsening toward consensus.

\end{abstract}


\maketitle

\section{Introduction}

A simple description for consensus in an interacting population is based on
endowing each individual, or agent, with discrete opinion states that evolve
by kinetics in which neighboring agents tend to agree.  Perhaps the best
studied such example is the 2-state voter model \cite{voter}, in which a
randomly-selected agent adopts the state of one of its neighbors.  This
update is repeated until a finite system necessarily reaches consensus.  When
the densities of agents of each state are equal, the time to reach consensus
in $d$ dimensions scales linearly in the number of agents $N$ for $d>2$, as
$N\ln N$ for $d=2$, and as $N^{2/d}$ for $d<2$ \cite{voter,pk}.  Another
classical description for consensus formation is the Ising model with
zero-temperature Glauber kinetics \cite{glauber}.  Here, a randomly-picked
agent adopts the state of the majority in its local interaction neighborhood;
in case of a tie, the initial agent flips with probability 1/2.  While this
updating promotes agreement, consensus does not necessarily arise.  In fact,
the system almost always gets stuck in infinitely long-lived metastable
states for spatial dimension $d\geq 3$ \cite{SKR}.

Recently, another prototypical model for consensus formation, majority rule
(MR), was introduced \cite{galam}.  In MR, a system consists of 2-state
agents, corresponding to two distinct opinions.  Agents evolve by the
following two steps: first, pick a group of agents with fixed odd size $G$;
this group is an arbitrary set of agents in the mean-field limit, or a
contiguous group for finite spatial dimension.  Second, all agents in this
group adopt the local majority state.  These two steps are repeated until
consensus is reached \cite{MM,MR2D}.  The MR model is analogous to the
majority rule of the Ising model with zero-temperature Glauber kinetics,
except that more than one agent can flip in a single update step.  A related
multi-spin kinetics occurs in the Sznajd model \cite{szn}, where a small
contiguous group of same-species spins induces all spins on the group
boundary to flip.  A perhaps desirable feature of the MR model is that the
group-based kinetics ensures that consensus is always reached in a finite
system.

\begin{figure}[ht] 
  \vspace*{0.cm}
  \includegraphics*[width=0.25\textwidth]{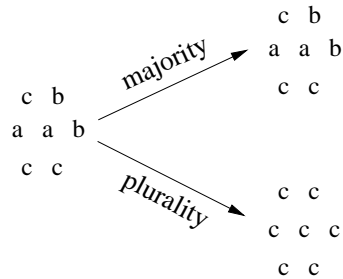}
  \caption{Illustration of a single update in majority and in plurality rule
    for a $3$-state system with group size $G=7$.}
  \label{process}
\end{figure}

In this work, we study two natural extensions of majority rule
(Fig.~\ref{process}).  The first is to allow each agent to have more than 2
equivalent states, {\it i.e.}, each agent possesses a Potts-like opinion
variable that may be in any of $s$ states $a,b,c,\ldots,s$ \cite{GGP}.  The
update step is similar to that in the original MR model.  A group of $G$
agents, with $G$ an odd number, is first identified.  If a local majority in
this group exists, that is, if $(G+1)/2$ or more agents in the group are in a
single state, then all agents adopt this local majority state.  However, if
there is no local majority, then the group does not evolve.  A new feature of
the multi-state compared to the 2-state model is the possibility of static
groups in which there is no local consensus.  Because of this feature, a
finite system does not necessarily reach consensus.  One of our goals will be
to characterize the dynamics of MR with more than 2 states and to determine
whether the final outcome is consensus or a frozen state.

A second extension is plurality rule (PR), which is meaningful only if the
number of spin states is 3 or greater.  In a single update step of PR, a
group of size $G$ is first identified.  Next the state with the most
representatives in that group---the plurality state---is determined.  All
agents in this group then adopt this plurality state.  In case of a tie, the
agents all adopt one of the multiple plurality states equiprobably.  This
type of evolution mimics how consensus might be achieved in parliamentary
systems when there are more than two parties.  Clearly, the PR model avoids
the freezing phenomenon that can occur in the multistate MR model.  We will
investigate how the PR model approaches consensus and the basic differences
between MR and PR dynamics.

In Sec.~II, we study the MR and PR models in the mean-field limit and show
generically that the consensus time grows logarithmically with $N$.  Then in
Sec. III, we discuss the change from diffusive domain coarsening, when the
number of opinion states $s$ and the group size $G$ are small, to ballistic
domain evolution, and finally to no evolution as $s$ and $G$ increase.  In
Sec.~IV, we present simulation results about the consensus time distribution
for the multi-state MR model that illustrate these two regimes of behavior.
In Sec.~V, we discuss the dynamical behavior of the PR and compare with the
MR model.  We conclude in Sec.~VI.

\section{Mean-field limit}

\subsection{Two-State Model}

We begin with the mean-field solution for the 2-state MR model in the
continuum $N\to\infty$ limit.  While the mean-field solution for the
$2$-state MR model with a finite number of agents was obtained previously
\cite{MM}, the continuum solution is much simpler, while still exhibiting
nearly all of the features of the discrete solution.

For simplicity, consider first the case of group size $G=3$.  In an update
step, the number of agents in state $a$, $N_a$, increases by 1 if the group
state is $aab$, while this number of agents decreases by 1 if the group state
is $bba$.  Thus $dN_a\!=3(a^2b-ab^2)$, where $a$ and $b$ now denote the
global densities of agents of each type, and the factor of 3 accounts for the
3 permutations of agents in the group.  If there is local consensus in a
group, then $N_a$ does not change in an update.  We increment the time by
$dt=3/N$ in an update so that each agent typically flips once in a single
time unit.  With these preliminaries, the rate equations for the agent
densities are
\begin{eqnarray}
\label{rate2}
\dot a&=&a^{2}b-b^{2}a \nonumber \\
\dot b&=&b^{2}a-a^{2}b. 
\end{eqnarray}

Since the total density $a+b=1$, we rewrite Eq.~(\ref{rate2}) as
\begin{equation}
\label{rate-1}
\dot a = ab(a-b)= -2 a \left(a-1/2\right)(a-1).
\end{equation}
In the latter form, we see that $a=0, 1$ are stable fixed points, while
$a=1/2$ is unstable.  Thus starting from any $a\neq 1/2$, the system is
quickly driven to consensus because of the non-zero bias inherent in
Eq.~(\ref{rate-1}).  The existence of a bias contrasts with the voter model,
where the average magnetization is conserved \cite{voter}, and purely
diffusive dynamics governs the evolution.

To determine the time until consensus is reached, we rewrite
Eq.~(\ref{rate-1}) in the partial fraction expansion
\begin{equation}
\label{da}
\left[\frac{1}{a}+\frac{1}{a-1}-\frac{2}{a-\frac{1}{2}}\right]\, da=-dt
\end{equation}
and integrate this equation of motion between suitable initial and final
states.  To describe a finite system of $N$ agents, we chose
$a_0=\frac{1}{2}+\frac{1}{N}$ and $a_\infty = 1-\frac{1}{N}$, corresponding
to the initial densities of the two species being equal and a final state of
consensus.  Integrating Eq.~(\ref{da}) over this range, we obtain
\begin{eqnarray}
  t\sim 3\ln N.
\end{eqnarray}

Thus, as previously found in the exact discrete solution \cite{MM}, the
consensus time $t$ scales as $\ln N$.  This dependence is a consequence of
the driving force in Eq.~(\ref{rate-1}) vanishing linearly as a function of
the distance to the fixed points $a=0, 1/2$, and $1$.  However, the
dependence of the amplitude in the consensus time is different in the
discrete and continuum solutions.  In the former case, the amplitude suddenly
drops from 2 to 1 as $|a-1/2|$ becomes larger than ${\cal O}(1/\sqrt{N})$
\cite{MM}, while in the continuum solution, the amplitude changes from 3
to 1 as $a-1/2$ becomes comparable to either $a$ or $1-a$.

The above considerations can be extended to larger group sizes.  By laborious
enumeration of all states up to $G=11$, it appears that the extension of
Eq.~(\ref{rate-1}) to general group size is
\begin{equation}
\label{rate-G}
\dot a = ab(a-b)\left[1 + 3ab+ 10 (ab)^2+ \ldots + 
\left(\!\!{j\atop{\frac{j+1}{2}}}\!\!\right)(ab)^{(j-1)/2}\right],
\end{equation}
where $j=G-2$.  Using Stirling's approximation, the higher-order terms are
all ${\cal O}(1/\sqrt{G})$ and there are ${\cal O}(G)$ such terms.  Thus
near the unstable fixed point, namely $a=\frac{1}{2}(1+\epsilon)$, the
equation of motion reduces to
\begin{equation}
  \dot \epsilon \propto \frac{\epsilon}{2} \sqrt{G}.
\end{equation}
Thus for general group size $G\ll N$, the time to reach consensus should
scale as $(\ln N)/\sqrt{G}$.

\subsection{More Than Two States}

We now extend our considerations to the multi-state MR model.  Consider first
the simplest case of 3 states ($s=3$) and group size $G=3$.  In the same
spirit as Eq.~(\ref{rate-1}), the rate equations for this 3-state model are
\begin{eqnarray}
\label{rate3}
\dot a&=&a^{2}(b+c)-a(b^{2}+c^{2}) \nonumber \\
\dot b&=&b^{2}(c+a)-b(c^{2}+a^{2}) \nonumber \\
\dot c&=&c^{2}(a+b)-c(a^{2}+b^{2}), 
\end{eqnarray}
with the densities of these states subject to the normalization condition
$a+b+c=1$.

\begin{figure}[!ht] 
  \vspace*{0.cm}
  \includegraphics*[width=0.4\textwidth]{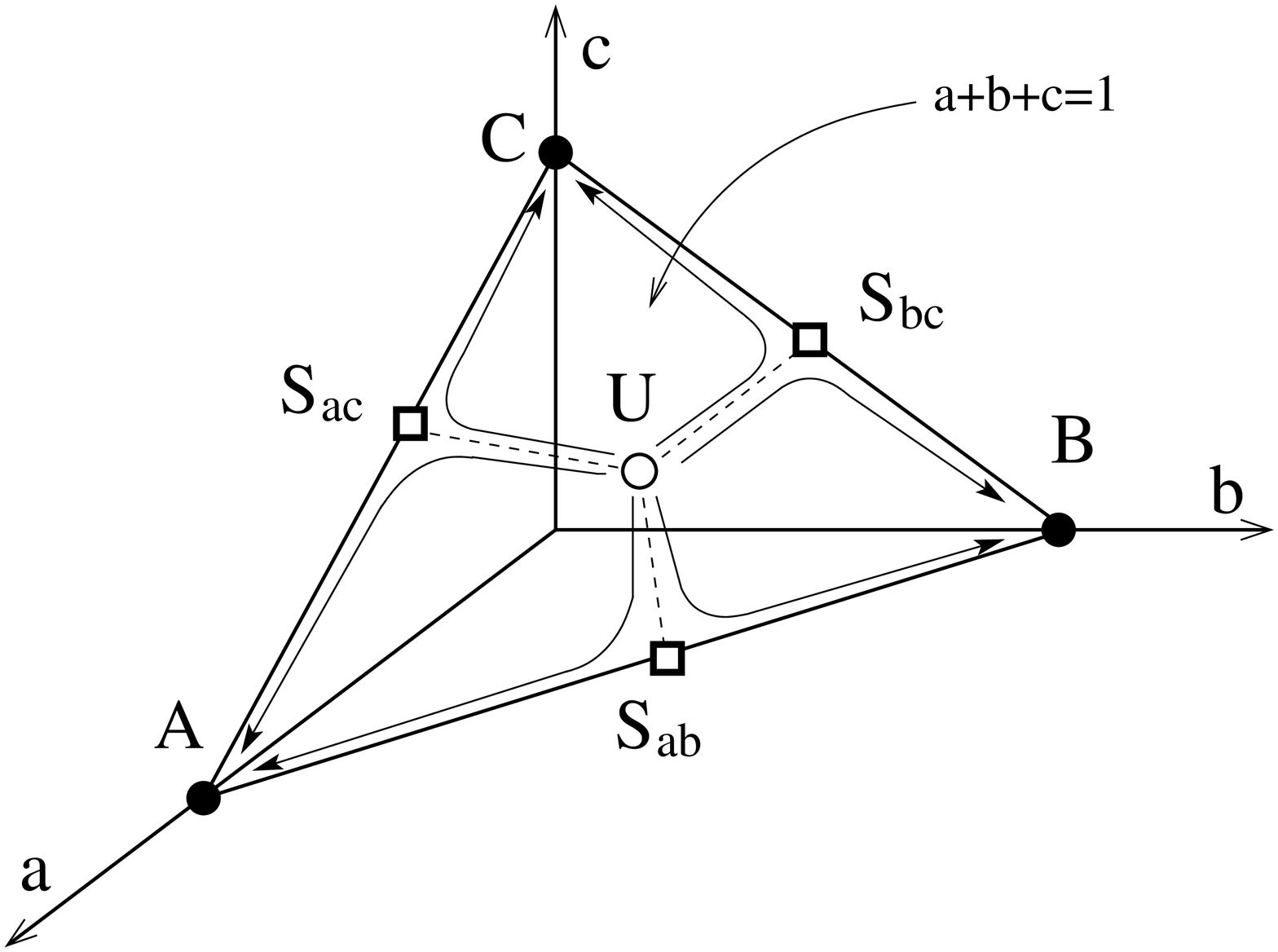}
  \caption{The phase plane $a+b+c=1$ in density space, showing the unstable fixed
    point ($\circ$), saddle points ($\square$), and stable fixed points
    ($\bullet$).  Separatrices are shown dashed.  Typical flows near the
    separatrices are also shown.}
  \label{triangle}
\end{figure}

To understand the resulting dynamics, we examine the stability of the fixed
points in these rate equations.  There are 7 such points; a globally unstable
fixed point at $U\equiv (\frac{1}{3},\frac{1}{3},\frac{1}{3})$, three saddle
points $S_{ab}\equiv (\frac{1}{2},\frac{1}{2},0)$, $S_{ac}\equiv
(\frac{1}{2},0,\frac{1}{2})$, and $S_{bc}\equiv (0,\frac{1}{2},\frac{1}{2})$,
and 3 stable points $A\equiv (1,0,0)$, $B\equiv (0,1,0)$, and $C\equiv
(0,0,1)$ (Fig.~\ref{triangle}).  There are 3 separatrices that join the
unstable fixed point to each of the saddle points (Fig.~\ref{triangle}).

We now compute the time to reach consensus starting from the unstable fixed
point.  There are two natural choices for the path to consensus.  One is to
run from the unstable fixed point to a saddle point close to a separatrix,
and then flow to the stable fixed point, as shown in Fig.~\ref{triangle}.
This path, $(\frac{1}{3},\frac{1}{3},\frac{1}{3})\to
(\frac{1}{2},\frac{1}{2},0)\to (1,0,0)$, corresponds to one species
disappearing first, while the other two have the same density, before
ultimate consensus is reached.  The other is the direct route
$(\frac{1}{3},\frac{1}{3},\frac{1}{3})\to (1,0,0)$ along a path where $b=c$.
That is, $b$ and $c$ both disappear at the same rate as consensus is reached.

For the indirect route, we first calculate the time required to go from $U$
to $S_{ab}$.  With the conditions $a=b$ and $c=1-2a$, the first of
Eqs.~(\ref{rate3}) can be rewritten as
\begin{eqnarray}
\dot a&=&a^{2}[a+(1-2a)]-a[a^{2}+(1-2a)^{2}] \nonumber \\
&=&-6a(a-1/2)(a-1/3).
\end{eqnarray}
Thus the time until the $c$ density vanishes is
\begin{eqnarray}
\label{mr-tc}
t_c&\approx&\int^{\frac{1}{2}-\frac{1}{N}}_{\frac{1}{3}+\frac{1}{N}}
\left[-\frac{1}{a}-\frac{2}{a-\frac{1}{2}}+\frac{3}{a-\frac{1}{3}}\right]\,da
\nonumber \\ 
&=&-2\ln(6/N)+3\ln(N/6)-6\ln(3/2)
\nonumber \\
&\approx&5\ln N.
\end{eqnarray}
Once again, the integration limits are chosen to correspond to the system
being infinitesimally close to, but not exactly at the fixed points at either
end of the trajectory.  Similarly, the time to go from $S_{ab}$ to $A$
asymptotically scales as $3\ln N$.  Thus the total time to go from
$(\frac{1}{3},\frac{1}{3},\frac{1}{3})$ to consensus asymptotically scales as
$8\ln N$.  Following the same line of reasoning, the consensus time 
from $(\frac{1}{3},\frac{1}{3},\frac{1}{3})$ to $(1,0,0)$ directly along the
path where $b= c$ always, asymptotically scales as $4\ln N$.

For the case $G=3$, we may generalize to obtain the consensus time for an
arbitrary number of states $s$.  The rate equations (\ref{rate3}) become
\begin{equation}
\label{rate-s}
\dot a_1=a_1^{2}(a_2+a_3+\cdots +a_s)-a_1(a_2^2+a_3^2+\cdots +a_s^2),
\end{equation}
plus cyclic permutations.  Here $a_k$ now denotes the density of the $k^{\rm
  th}$ species.  Again, we compute the consensus time along a sequential path
in which one species disappears first while the remaining species have equal
densities, as well as along a direct path in which $s-1$ species have equal
densities that all go to zero simultaneously.

For the first leg of the sequential path, the densities satisfy the
conditions $a_1=a_2=a_3=\cdots =a_{s-1}=a$ and $a_s=1-(s-1)a$.  Then
Eq.~(\ref{rate-s}) reduces to
\begin{equation}
 \label{rate-s1}
\dot a=-s(s-1)\,\,a\left(a-\frac{1}{s-1}\right)\left(a-\frac{1}{s}\right).
\end{equation}
Integrating from the unstable fixed point to the fixed point where the
density of one species is zero, the transit time equals $(2s-1)\ln N$.
Repeating this calculation as each species is eliminated, the consensus time
is
\begin{equation}
t=\sum_{k=2}^{s} (2k-1)\ln N =(s^2-1)\ln N
\end{equation}
Similarly, for consensus via a direct path, the densities satisfy the
constraints $a_1=a$ and $a_2=a_3=\cdots =a_s=(1-a)/(s-1)$.  Now
Eq.~(\ref{rate-s}) becomes
\begin{equation}
  \label{rate-s2}
\dot a=-\frac{s}{s-1}\,\,a\left(a-\frac{1}{s}\right)\left(a-1\right),
\end{equation}
from which the consensus time equals $(s+1)\ln N$.

In summary, the consensus time always scales as $\ln N$, with a prefactor
that is an increasing function of the number of states.  This coefficient
depends on the actual route to consensus.  If species are eliminated one by
one, the coefficient grows quadratically with $s$, while if one species
dominates and the remaining $s-1$ all disappear at the same rate, the
coefficient grows linearly with $s$.  Qualitatively similar results arise for
larger group sizes.

\subsection{Plurality Rule}

Finally, we study plurality rule (PR) which first becomes distinct from MR
when number of states $s \ge 3$ and the group size $G \ge 5$.  Let us study
this case $s=3, G=5$ for simplicity.  By enumerating all group states, the
rate equations for the PR model are:
\begin{eqnarray}
\label{rate-mr}
\begin{split}
\dot a&\!=\!a^{4}(b\!+\!c)\!+\!4a^3(b\!+\!c)^2\!-\!4a^2(b^3\!+\!c^3)\!-\!4abc(b^2\!+\!c^2)\\
&~~~~-a(b^4+c^4)+3a^2bc(b+c)-6ab^2c^2, \\
\end{split}
\end{eqnarray}
plus cyclic permutations for $\dot b$ and $\dot c$.  The rate equations for
the MR model for $s=3$ and $G=5$ are identical to Eq.~(\ref{rate-mr}) except
that the last two terms, corresponding to states with no global majority, are
absent.

The MR and PR models both share the same global fixed point structure and
stability.  Let us now determine the transit time from
$(\frac{1}{3},\frac{1}{3},\frac{1}{3})\to (\frac{1}{2},\frac{1}{2},0)$ along
the path $a=b$ for the case $s=3, G=5$ for both the PR and MR models.  Using
$a=b$ and $c=1-2a$, the rate equation for $\dot a$ reduces to
\begin{eqnarray}
  \label{MP}
\dot a&=&48a^5-66a^4+25a^3-a\quad  ({\rm MR}) \nonumber \\
\dot a&=&36a^5-60a^4+25a^3-a\quad  ({\rm PR})
\end{eqnarray}
To determine the transit time simply, note that the driving flow in the rate
equation vanishes linearly near the fixed points and that the coefficient of
the linear dependence determines the coefficient of $\ln N$ in the transit
time.  We thereby find $t\sim 5\ln N$ for PR, while $t\sim \frac{43}{5}\ln N$
for MR.  As expected, PR evolves faster than MR because there are fewer
possibilities for frozen groups.  Finally, to reach consensus via
$(\frac{1}{2},\frac{1}{2},0) \to (1,0,0)$, the rate equations for both the PR
and MR models are $\dot a=6a^5-15a^4+10a^3-a$, so that the time to go from
$S_{ab}$ to $A$ is $\frac{23}{7}\ln N$.  Thus the consensus time from $U$ is
$\frac{58}{7}\ln N$ for PR and $\frac{86}{7}\ln N$ for MR.  Thus there is
only a quantitative difference between the PR and MR models in the mean-field
limit.

\section{Diffusive vs.\ Deterministic Consensus}

\begin{figure}[ht] 
 \vspace*{0.cm}
  \includegraphics*[width=0.22\textwidth]{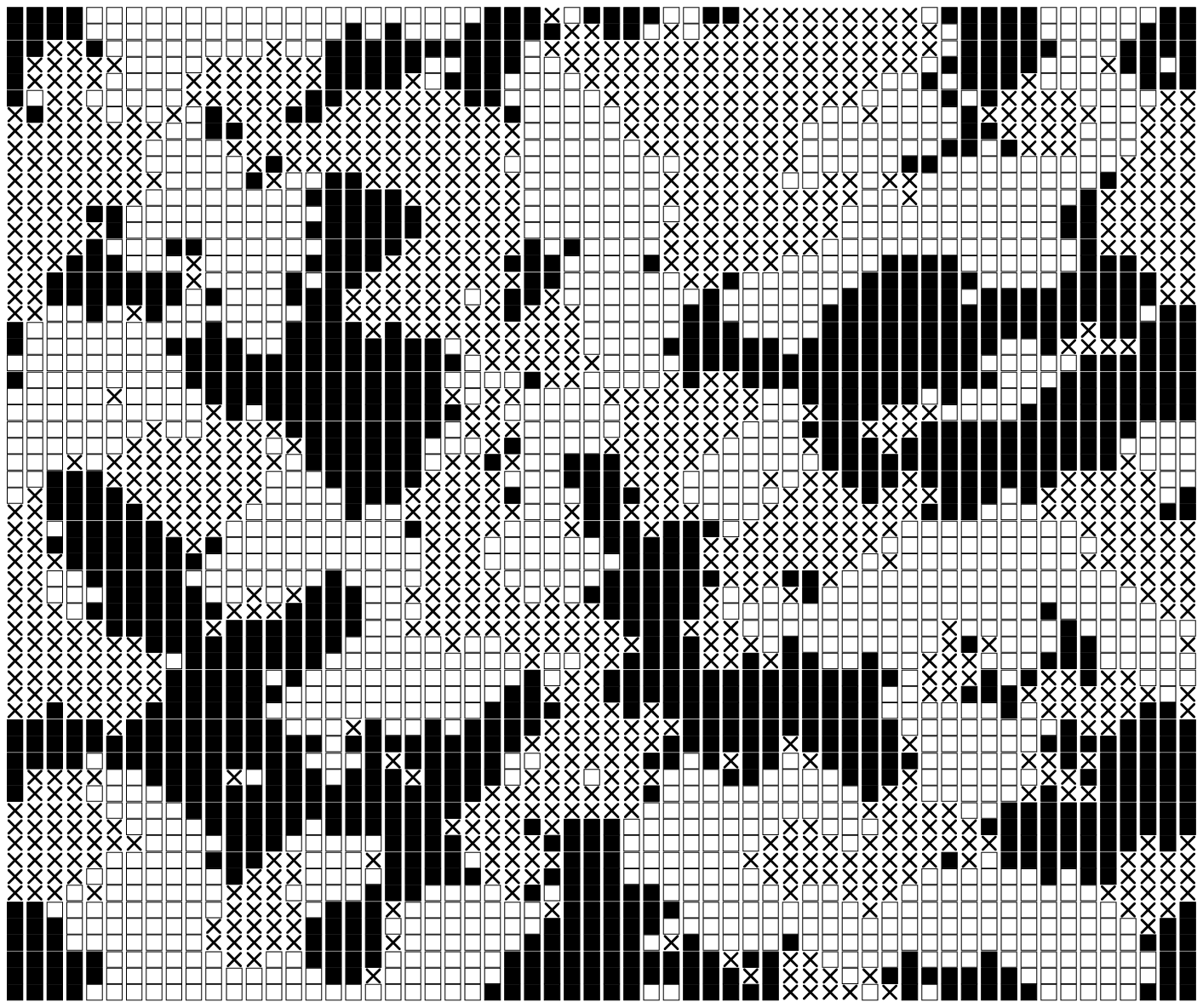}\hfil
  \includegraphics*[width=0.22\textwidth]{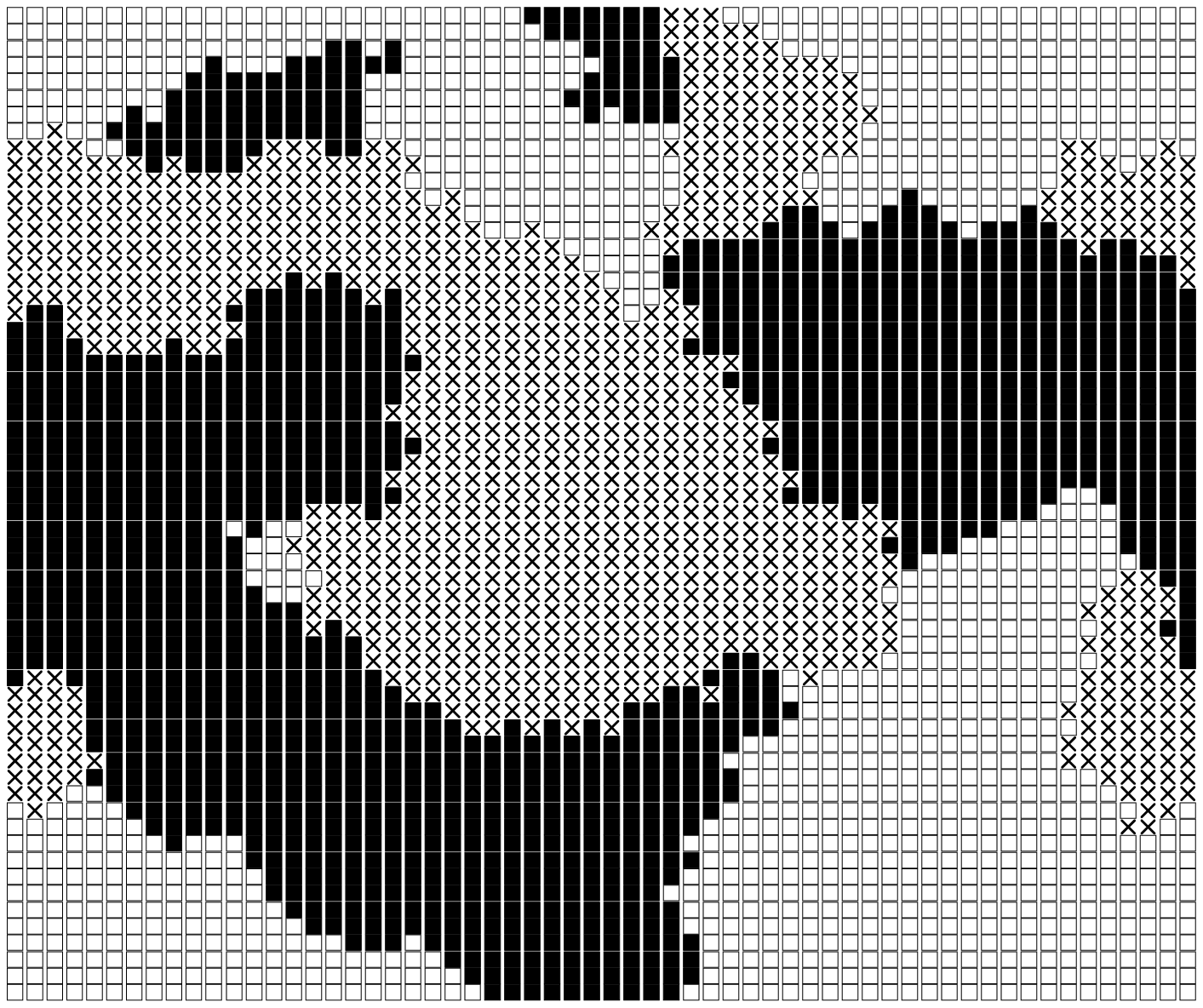}\\ \vskip 2ex
  \includegraphics*[width=0.22\textwidth]{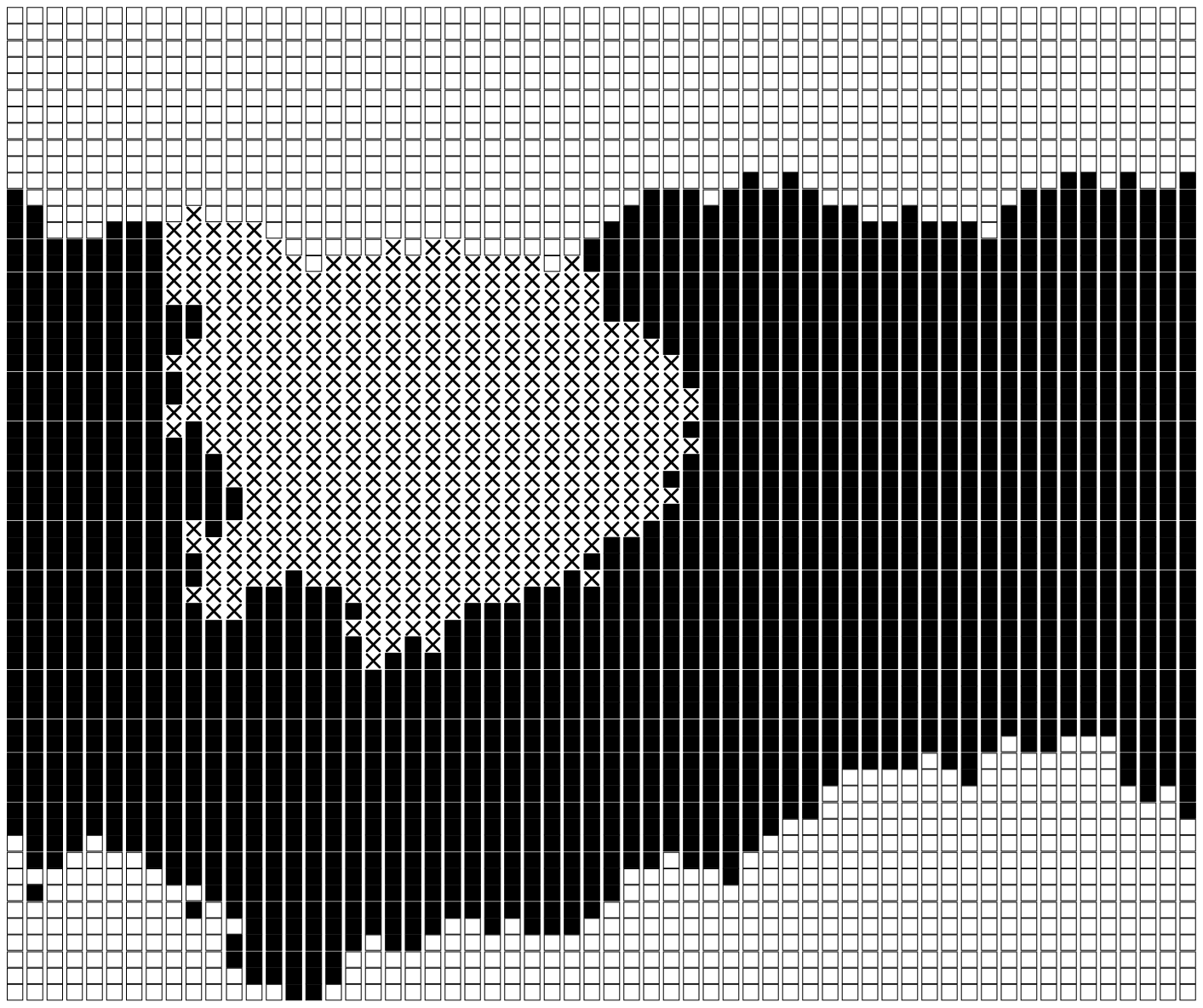}\hfil
  \includegraphics*[width=0.22\textwidth]{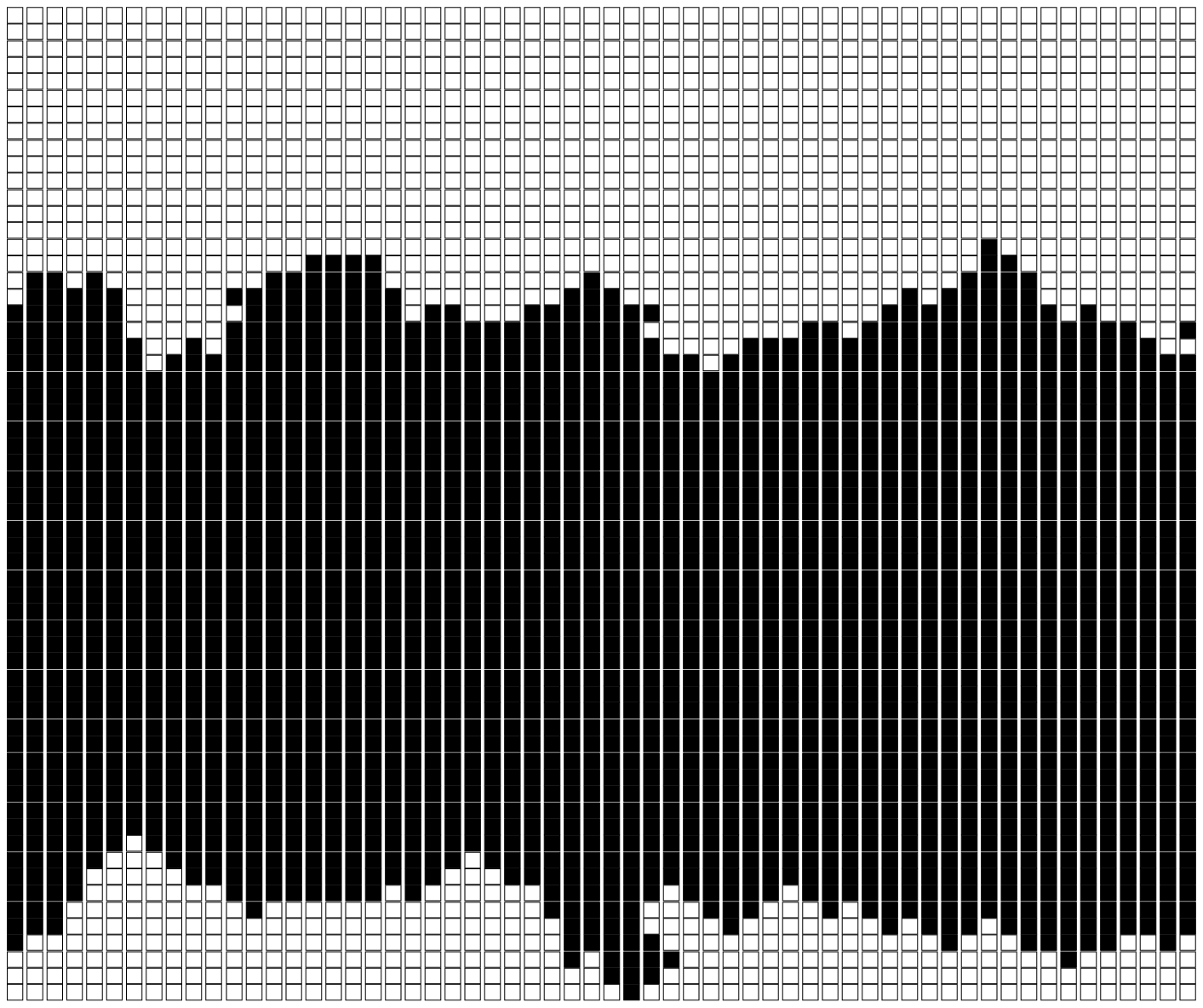}\\ \vskip 2ex
\vspace*{0.2cm}
\caption{Evolution of a $60\times 60$ 3-state MR model with group size $G=3$ at
  times $t=1,20,80$, and 640.}
\label{pics1}
\end{figure}

\begin{figure}[ht] 
 \vspace*{0.cm}
  \includegraphics*[width=0.22\textwidth]{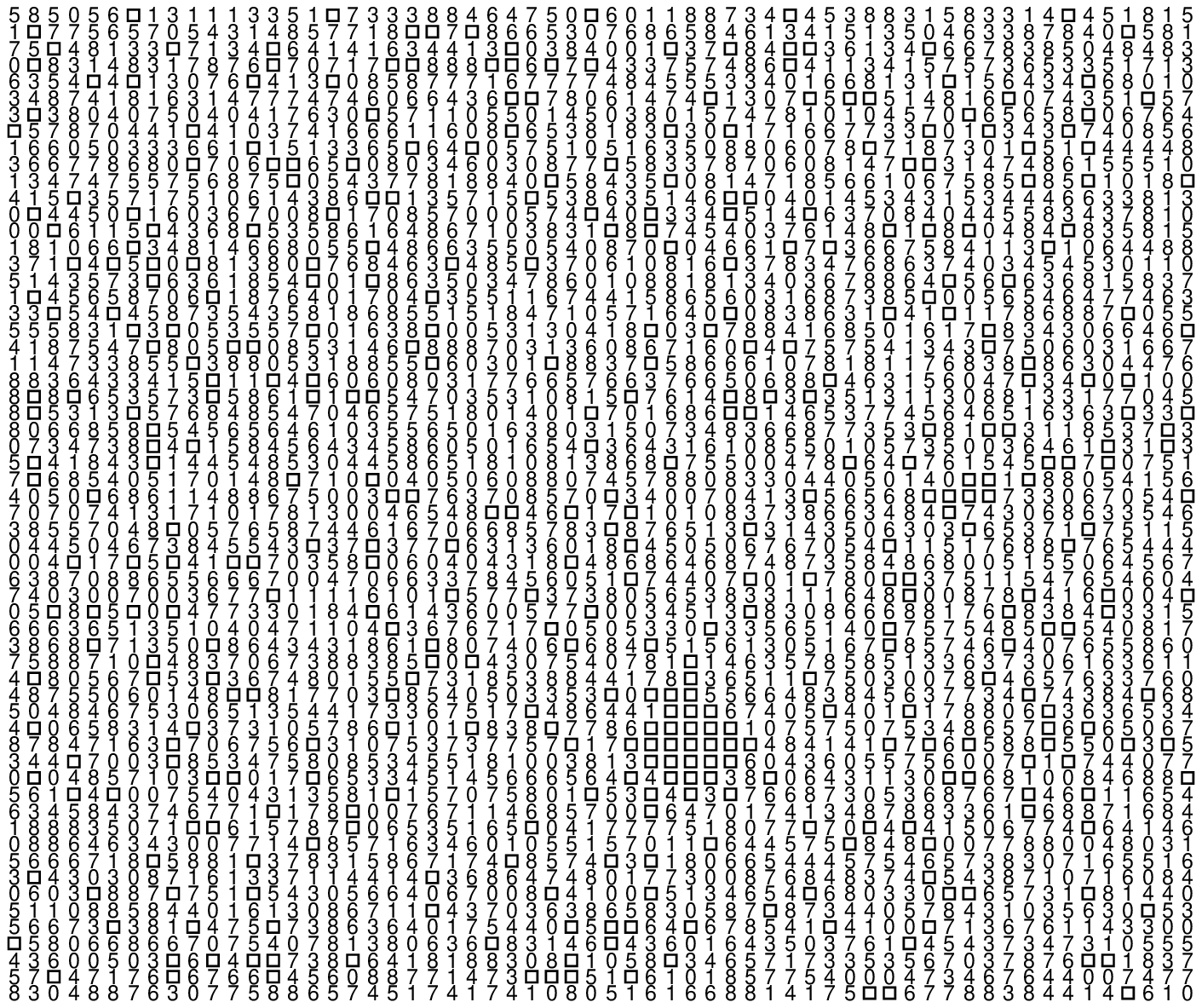}\hfil
  \includegraphics*[width=0.22\textwidth]{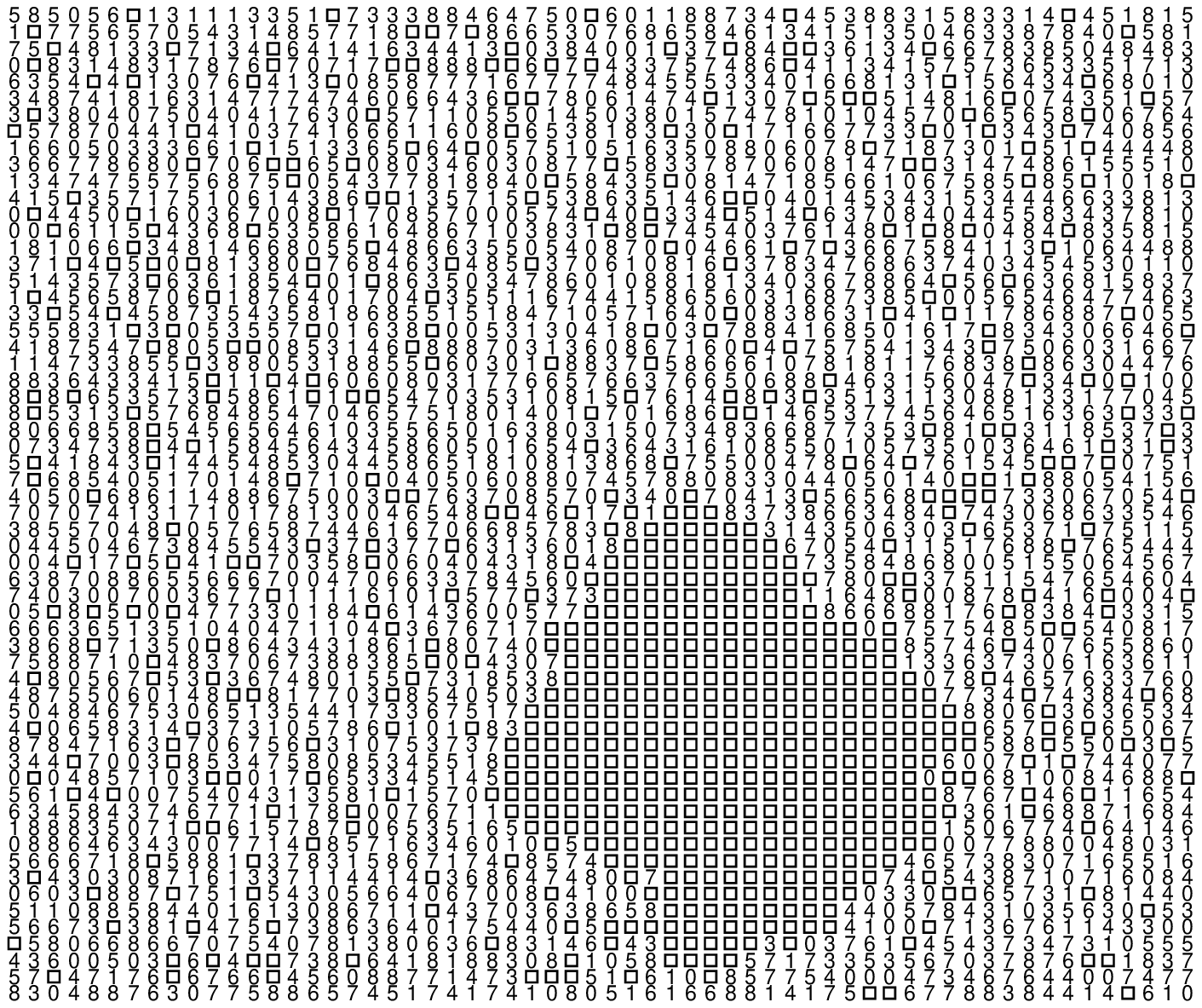}\\ \vskip 2ex
  \includegraphics*[width=0.22\textwidth]{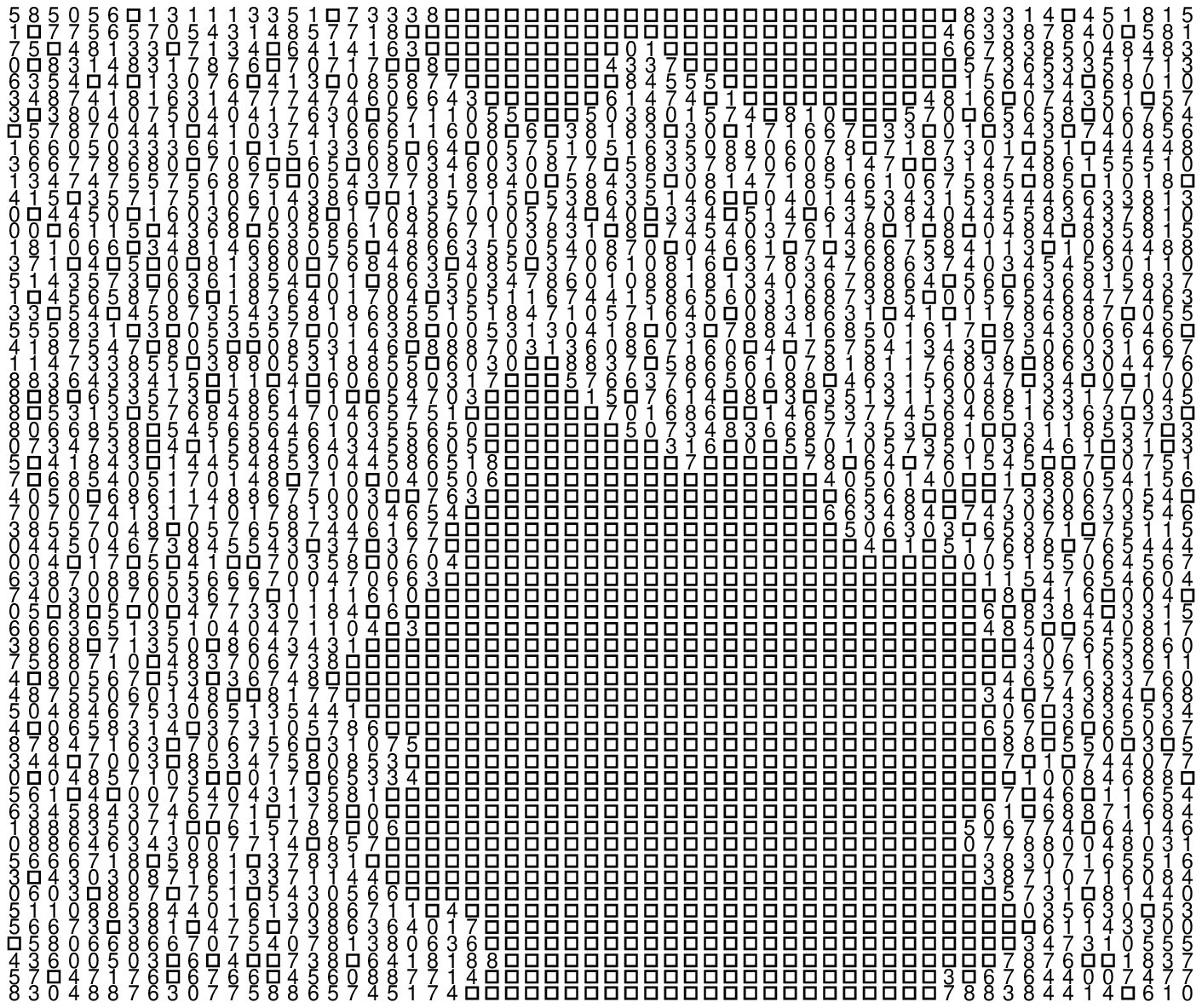}\hfil
  \includegraphics*[width=0.22\textwidth]{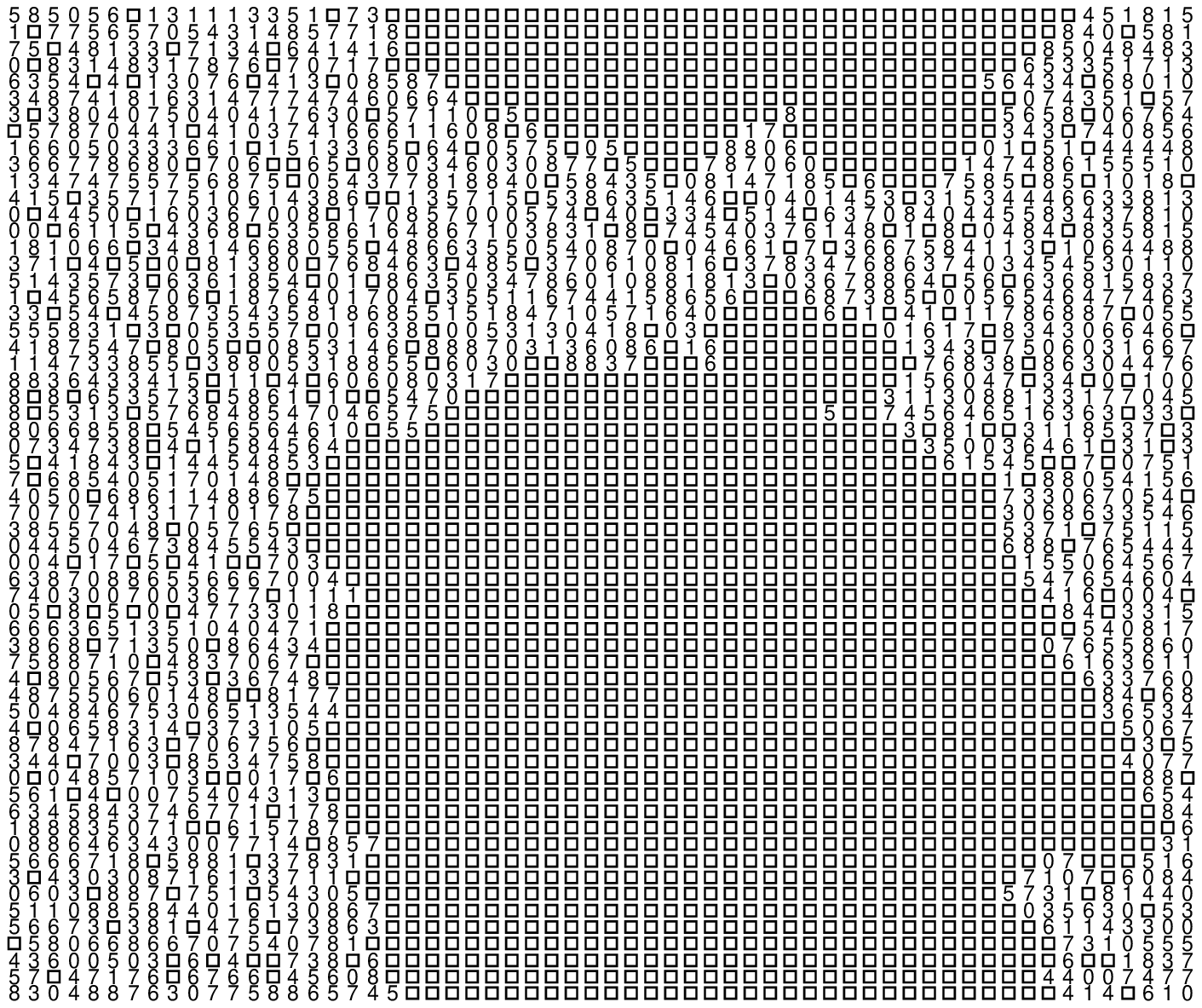}\\ \vskip 2ex
\vspace*{0.2cm}
\caption{Evolution of a $60\times 60$ 9-state MR model with
  group size $G=25$ at times $t=0.1,1.5,1.8$, and 2}
\label{pics2}
\end{figure}

We now investigate the multi-state MR model on the square lattice.  While the
rate equations predict that the number of states and the group size do not
qualitatively affect the evolution, simulations on the square lattice show
very different behaviors for systems with small $s$ and/or $G$ and those with
large $s$ and $G$ (Figs.~\ref{pics1} \& \ref{pics2}).  In the former case, a
coarsening domain mosaic evolves diffusively.  However, when both $s$ and $G$
are large, it is improbable that a small system will contain even a single
group with an initial local majority, and such a system will be static.  At
the boundary between these two regimes, there will typically be a single
initial local majority group.  When this occurs, there is nearly
deterministic evolution in which this group grows ballistically and overruns
the system (Fig.~\ref{pics2}).

To delineate these two regimes, we determine the criterion for the existence
of at least one group with a local majority in the initial state.  As a
preliminary, we determine the probability ${\cal P}_>$ that a local majority
exists in a single group of size $G$, {\it i.e.}, $n>G/2$ of the agents in
the group all have the same opinion.  The probability for this event to occur
is
\begin{equation}
\label{p1}
{\cal P}_>=s\sum_{n>G/2}^G \frac{G!}{n!(G-n)!}\left(\frac{1}{s}\right)^n 
\left(1-\frac{1}{s}\right)^{G-n}.
\end{equation}
Using Stirling's approximation, this expression reduces to
\begin{equation}
\label{G-approx}
{\cal P}_>\approx 2^G\,\frac{(s-1)^{G/2}}{s^{G-1}}\,\sqrt{\frac{2}{\pi G}}
\int_0^\infty e^{-2x^2/G}\, e^{-x\ln(s-1)}\, dx,
\end{equation}
where $x=n-G/2$ and we extend the upper limit of $x$ from $G/2$ to $\infty$.

For $s=2$, Eq.~(\ref{G-approx}) gives ${\cal P}_>=1$, the obvious exact
result, suggesting that this Gaussian approximation may be relatively
accurate.  Conversely, for large $s$, the exponential in the integrand of
Eq.~(\ref{G-approx}) decays more rapidly than the Gaussian factor, and we
obtain
\begin{equation}
  \label{Ps}
  {\cal P}_>\approx \frac{s}{\sqrt{G} \ln s}\left(\frac{2}{\sqrt{s}}\right)^G
  \qquad s\to\infty.
\end{equation}
In fact, for $s=3$, the exponential cutoff controls the integral in
Eq.~(\ref{G-approx}) already for $G\approx 4$; for $s=4$, the exponential
cutoff dominates for $G\approx 2.23$.  Thus the approximate form
Eq.~(\ref{Ps}) is relatively accurate, except the special case $s=2$.
Clearly, the probability that a randomly-populated group of size $G$ has a
majority decreases rapidly when either $s$ or $G$ increases.

\begin{figure}[ht] 
  \vspace*{0.cm} 
\centerline{\includegraphics*[width=0.4\textwidth]{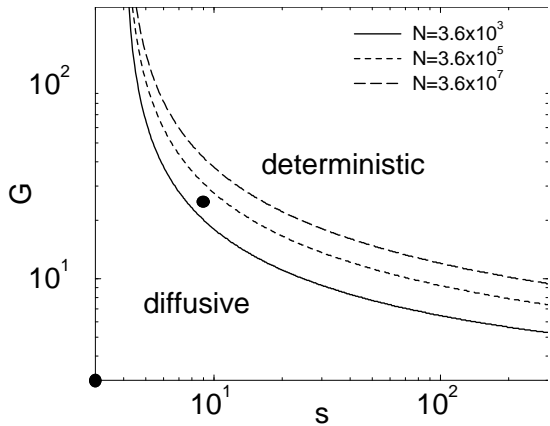}}
\caption{Phase diagram for the majority rule model as a function of the
  number of states $s$ and the group size $G$.  The curves are defined by the
  criterion $NP_> =1$, separates a region of diffusive dynamics (multiple
  initial majority groups) from a static region (no initial majority groups).
  The dots correspond to the two parameter values simulated in Sec.~IV.}
 \label{region}
\end{figure}

The condition on the number of agents $N$ for there to be at least one
initial group in the system that contains a local majority is thus $M{\cal
  P}_>>1$ (Fig.~\ref{region}), where $M$ is the number of independent groups
in the system.  While we do not know $M$ exactly, trivial bounds are $N/G\leq
M\leq N$.  The lower bound is based counting the $N/G$ contiguous groups that
tile the system as independent, while the latter is based on considering all
$N$ embeddings of the group on the lattice as independent.  Since ${\cal
  P}_>$ varies rapidly with $s$ and $G$, this small indeterminacy in $M$ does
not play a major role in the condition for the existence of a group with a
local majority.

From Eq.~(\ref{Ps}), the criterion $M{\cal P}_>>1$ becomes
\begin{equation}
  \label{MC}
M>M_c\approx \frac{\sqrt{G}\ln s}{s}\,e^{G\ln(\sqrt{s}/2)}.
\end{equation}
For $M>M_c$, the number of initial groups with a local majority is non-zero
so that evolution occurs.  If $M< M_c$, no groups have a local majority and
the system will therefore be static.  When $M\agt M_c$, majority groups are
widely separated and there is a two-stage approach to consensus.  First,
initial majority groups grow ballistically until they meet, after which
domain interfaces evolve diffusively toward final consensus.  The case
$M\approx M_c$ is exceptional as there is typically a single initial group
with a local majority that quickly overruns the system (Fig.~\ref{pics2}).
It is also noteworthy that for moderate values of $s$ and $G$, an
astronomically large system is needed for there to be at least one group with
a local majority.

\begin{figure}[ht] 
  \vspace*{0.cm} 
\centerline{\includegraphics*[width=0.4\textwidth]{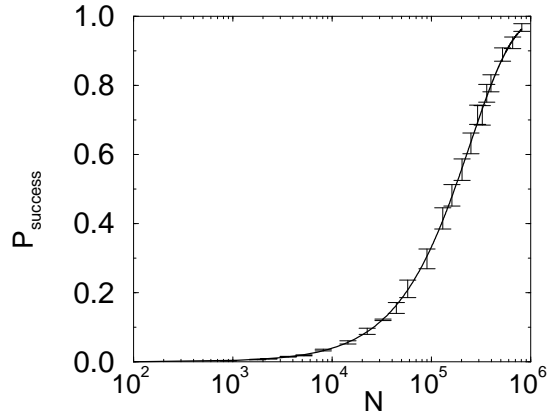}}
\caption{Probability for a successful realization versus system size for
  $s=9, G=25$ MR model based on using $M={N}/{8}$.  The error bars are based
  on a $5\%$ significance level.}
 \label{psvsN}
\end{figure}

To test Eq.~(\ref{MC}), we compare our analytical prediction for the
``success'' probability $P_{\rm success}=1-(1-P_>)^M$, defined as the
probability that a randomly-prepared configuration contains at least one
majority group, with corresponding simulation data from the $s=9, G=25$ MR
model (Fig.~\ref{psvsN}).  We observe that $P_{\rm success}$ increases
quickly when the system size passes through the threshold value $M_c$, and
that there is excellent agreement between the data and our analytics.

\section{Consensus Time Distribution in Two Dimensions}

We now study the evolution of the multi-state MR model on the square lattice
by numerical simulations.  We construct a group by incorporating successive
diamond-shaped shells with increasing values of $|x|+|y|$, where $x$ and $y$
are the horizontal and vertical distances from the central agent.  Thus for a
given $G$, the group is defined as the initial agent, plus the 4 agents in
the first shell, the 8 agents in the second shell, {\it etc.}, until a total
of $G$ agents are included.  To ensure that a group contains precisely $G$
agents, a randomly-selected set of agents in the last shell are typically
included.  We then evolve the system according to majority rule until
consensus is reached.

\begin{figure}[ht] 
 \vspace*{0.cm}
 \includegraphics*[width=0.4\textwidth]{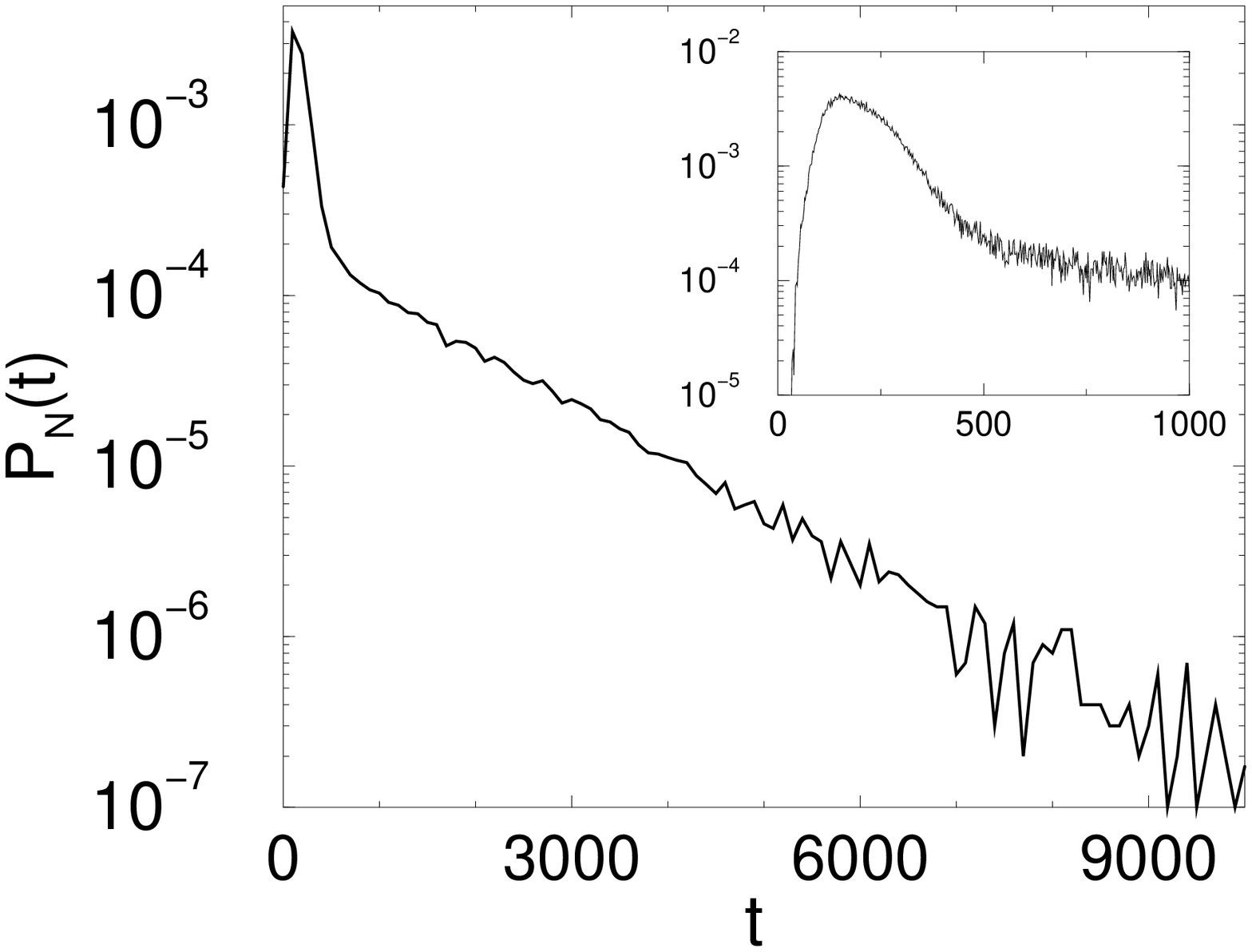}
 \includegraphics*[width=0.4\textwidth]{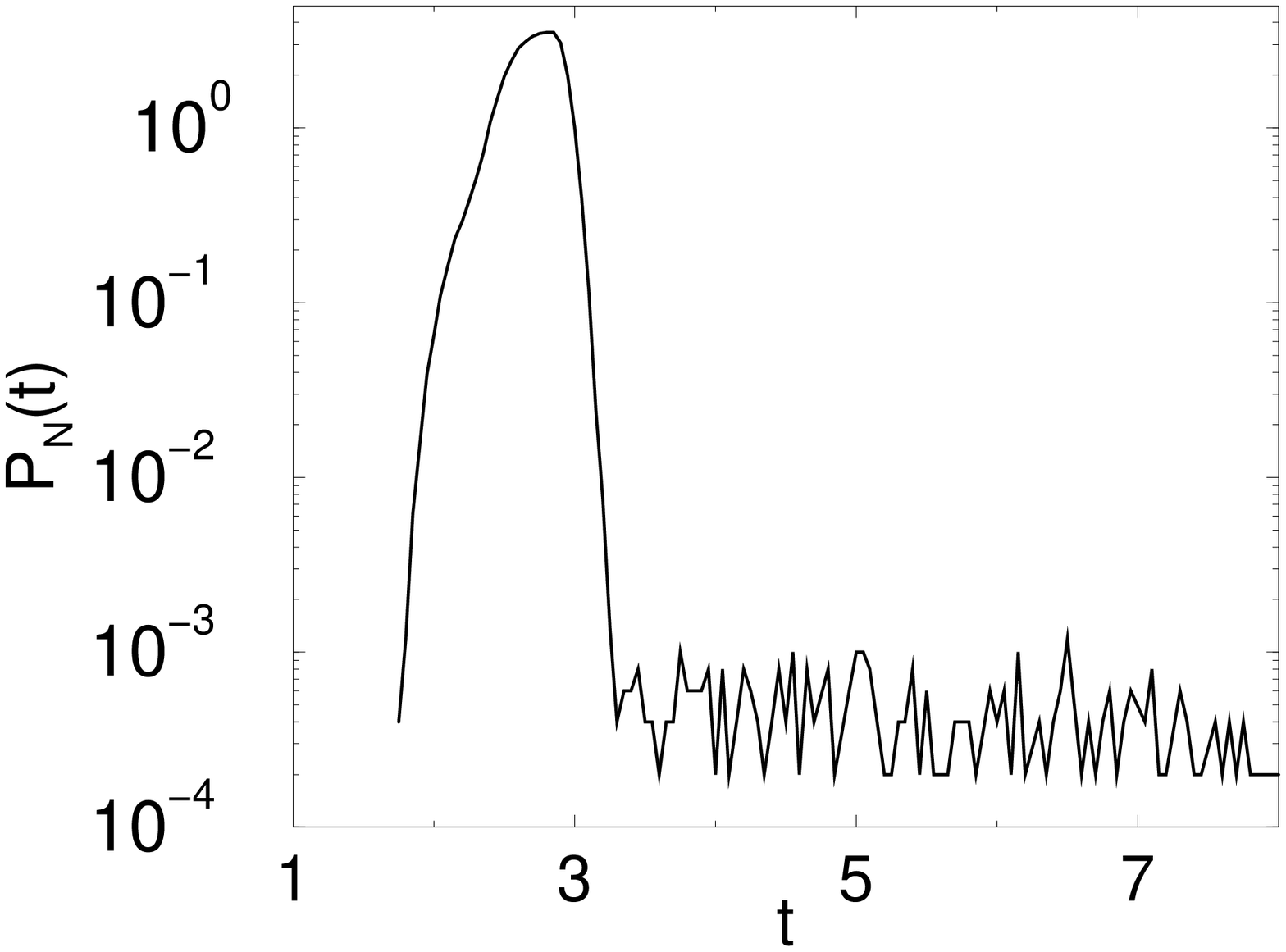}
 \caption{Probability distribution for the consensus time $P_N(t)$ versus $t$
   on a square lattice of $N=3600$ sites by majority rule for the two cases
   of $s=3, G=3$ with data integrated over bins of width $200$ (top), and
   $s=9, G=25$, and bin size $0.05$ (bottom). The inset shows detail near
   the peak where the width of the data bins is 2.}
 \label{pdf2}
\end{figure}

As in our previous study of the 2-state MR \cite{MR2D}, we focus on the
distribution of consensus times, $P_N(t)$.  To illustrate the behavior in the
two regimes of Fig.~\ref{region}, we consider the representative cases $s=3,
G=3$ and $s=9, G=25$, with $N=3600$ agents for both examples (heavy dots in
Fig.~\ref{region}), in which the initial concentrations of all species are
equal.  As shown in Fig.~\ref{pdf2}, there is a long-time tail in $P_N(t)$
for the $3$-state model that is qualitatively similar to that in the
$2$-state model.  For 3 states, approximately 1/5 of all configurations get
stuck in long-lived coherent stripe states, compared to approximately 1/3 of
all states in the $2$-state model \cite{MR2D}.  Analogous coherent states,
albeit with infinite lifetimes, also occur in the Ising model with
zero-temperature Glauber kinetics \cite{SKR}.

\begin{figure}[!ht] 
 \vspace*{0.cm}
 \includegraphics*[width=0.4\textwidth]{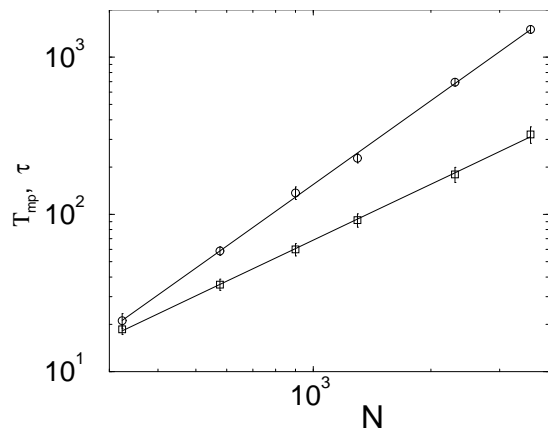}
 \caption{Dependence of the two characteristic times $T_{\rm mp}$ ($\Box$)
   and $\tau$ ($\circ$) on $N$ for the 3-state $G=3$ MR model.}
 \label{t-s3g3}
\end{figure}

For the 3-state model, we also measure the $N$-dependence of the two basic
time scales: (a) the most probable consensus time, $T_{\rm mp}\equiv
e^{\langle \ln t\rangle}$, corresponding to the location of the peak in
$P_N(t)$, and (b) the characteristic decay time in the exponential tail of
$P_N(t)\sim e^{-t/\tau}$.  As found previously for the 2-state model
\cite{MM,MR2D}, these time scales have power-law dependences on $N$ with
different exponents (Fig.~\ref{t-s3g3}).  By fitting the data to
straight lines, we find that $T_{\rm mp}$ varies as $N^{\alpha}$, where
$\alpha=1.18 \pm 0.03$, while $\tau$ varies as $N^{\beta}$, with $\beta
=1.77\pm 0.05$.  All quote error bars correspond to a $5\%$ significance
level.  These exponents are close to the corresponding values of $1.24$, and
$1.7$ respectively, quoted for the 2-state model \cite{MR2D}.  We conclude
that the same underlying domain wall diffusion mechanism governs the approach
to consensus in both the 2- and 3-state MR models.

\begin{figure}[!ht] 
 \vspace*{0.cm}
 \includegraphics*[width=0.4\textwidth]{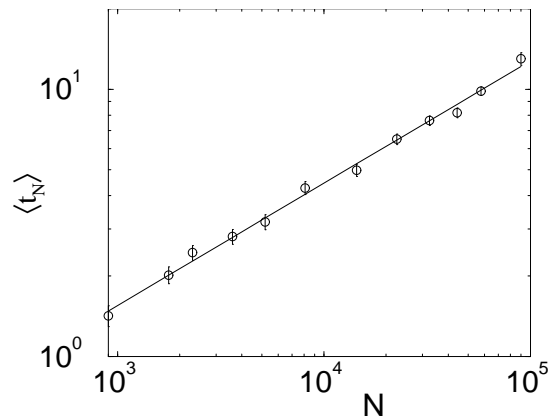}
 \caption{Dependence of the mean consensus time $\langle t_N\rangle$ versus
   $N$ for $9$-state $G=25$ MR model.}
 \label{t-s9g25}
\end{figure}

In contrast, for the $s=9, G=25$ MR model, the consensus time distribution is
sharply peaked about its most probable value (Fig.~\ref{pdf2}), so that there
is only a single characteristic time scale.  The absence of a second longer
time scale is due to the fact that long-lived configurations, such as stripe
states, no longer occur.  We find that the average consensus time grows only
as $N^\alpha$, with $\alpha=0.48\pm 0.04$ (Fig.~\ref{t-s9g25}).  Since
we are interested in the $N$-dependence of the consensus time for large $N$,
where almost all configurations are successful, we define this average only
for successful realizations.

We can justify the slow growth of the consensus time for the case $s=9, G=25$
by a simple-minded argument.  Since the evolution is determined by the growth
of a single domain, the consensus time is essentially the time for this
domain to overrun the system.  According to majority rule, a randomly
selected group typically evolves when more than half of its area is within
the dominating domain.  Thus after each time step, the radius of this domain
grows by an amount that is of the order of $G$.  When the domain radius
equals the linear dimension of the system, $Gt=\sqrt{N}$, and consensus is
reached.  Thus we obtain $t \approx {\sqrt{N}}/{G}$, which appears to account
for the $N$-dependence of the consensus time in Fig.~\ref{t-s9g25}.

\section{Majority vs.\ Plurality Rule}

\begin{figure}[H] 
 \vspace*{0.cm}
  \includegraphics*[width=0.22\textwidth]{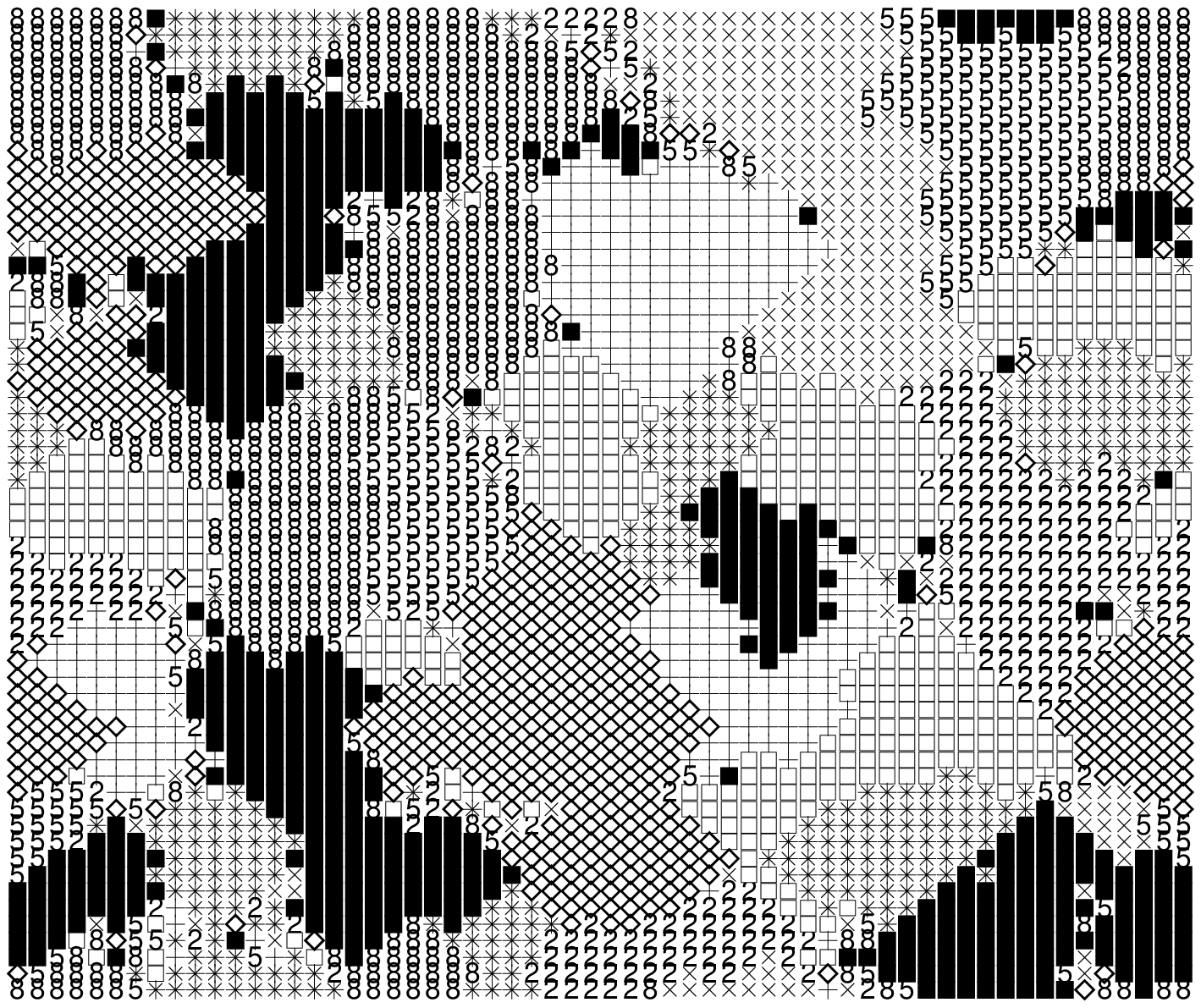}\hfil
  \includegraphics*[width=0.22\textwidth]{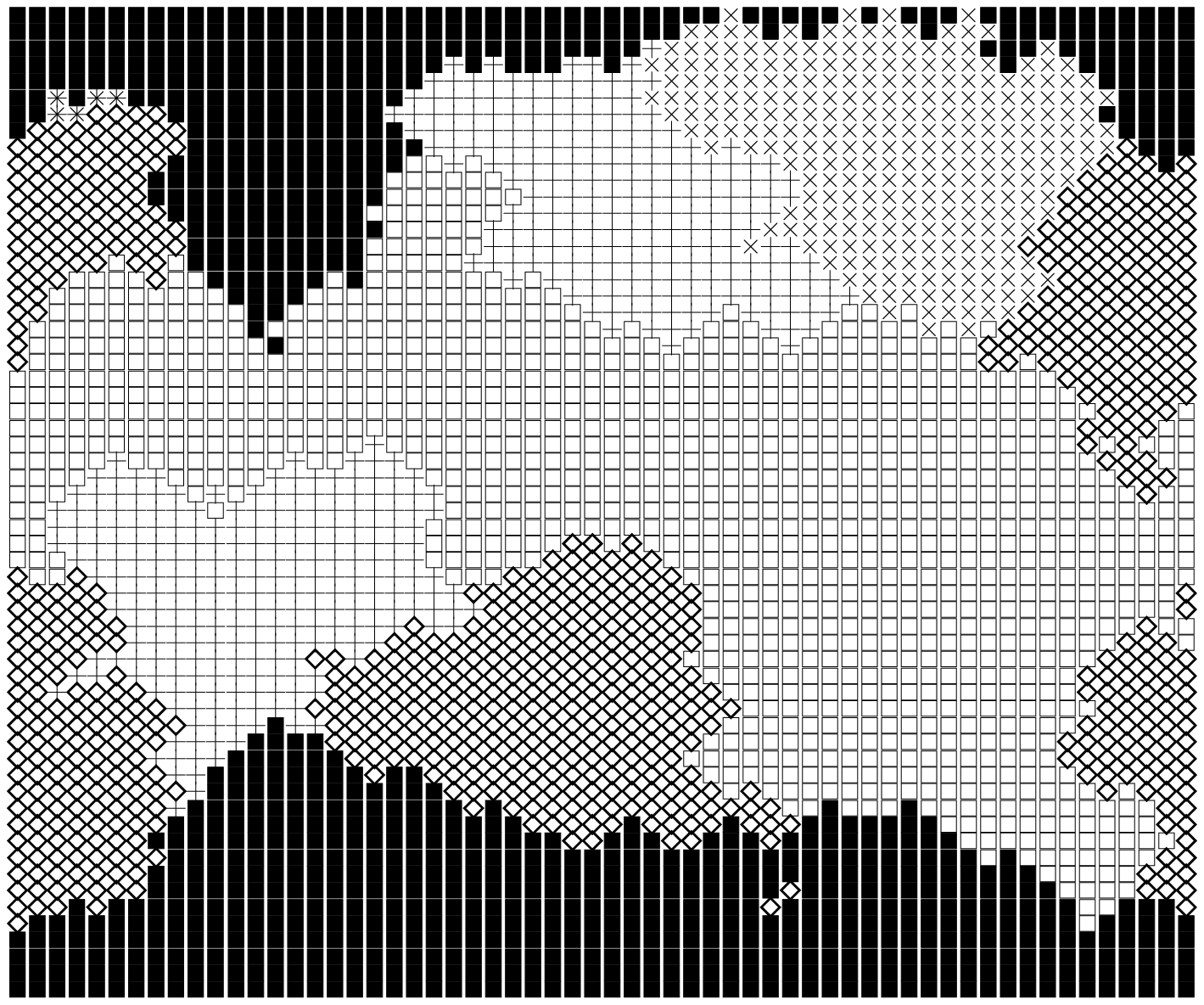}\\ \vskip -0.5ex
  \includegraphics*[width=0.22\textwidth]{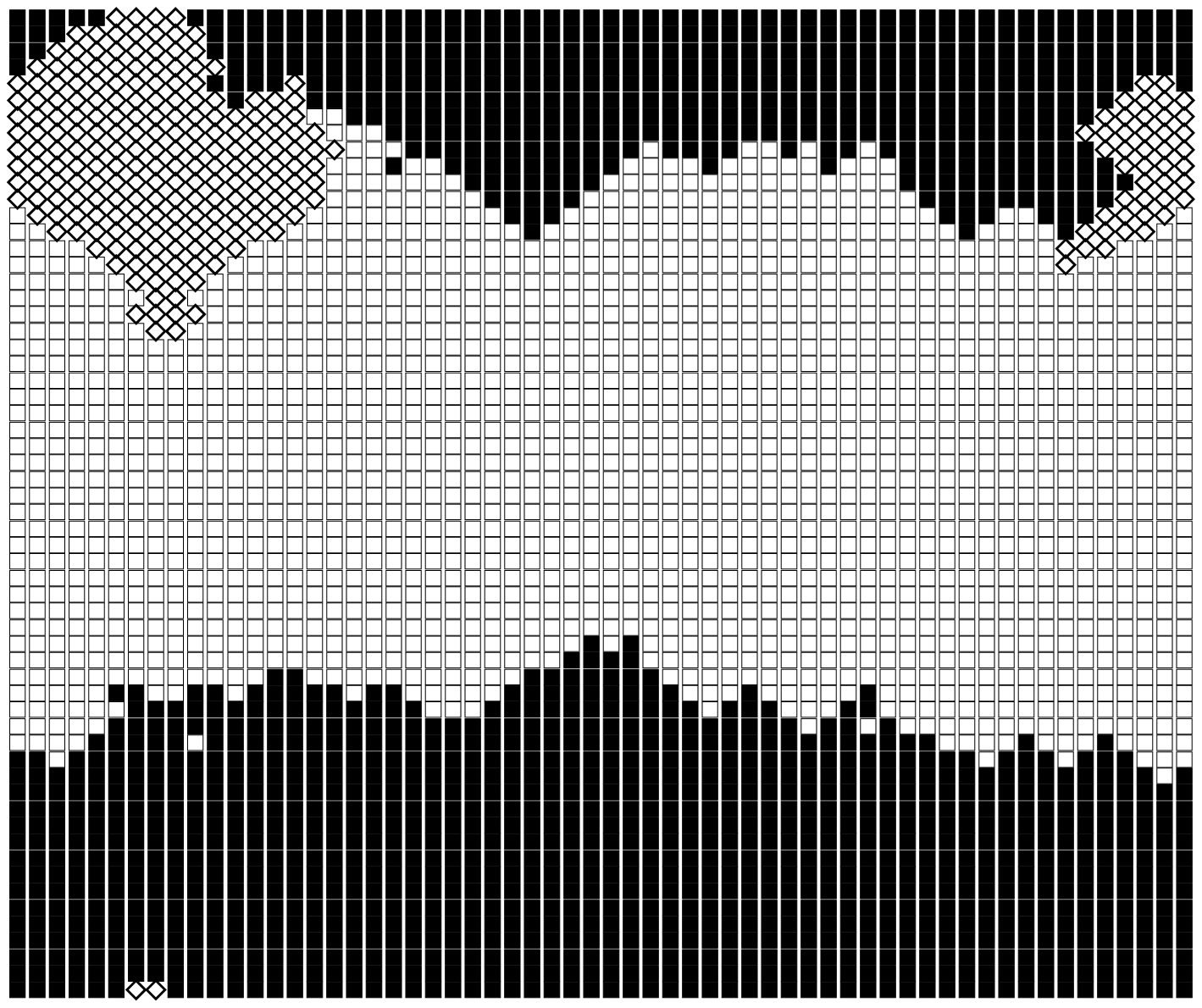}\hfil
  \includegraphics*[width=0.22\textwidth]{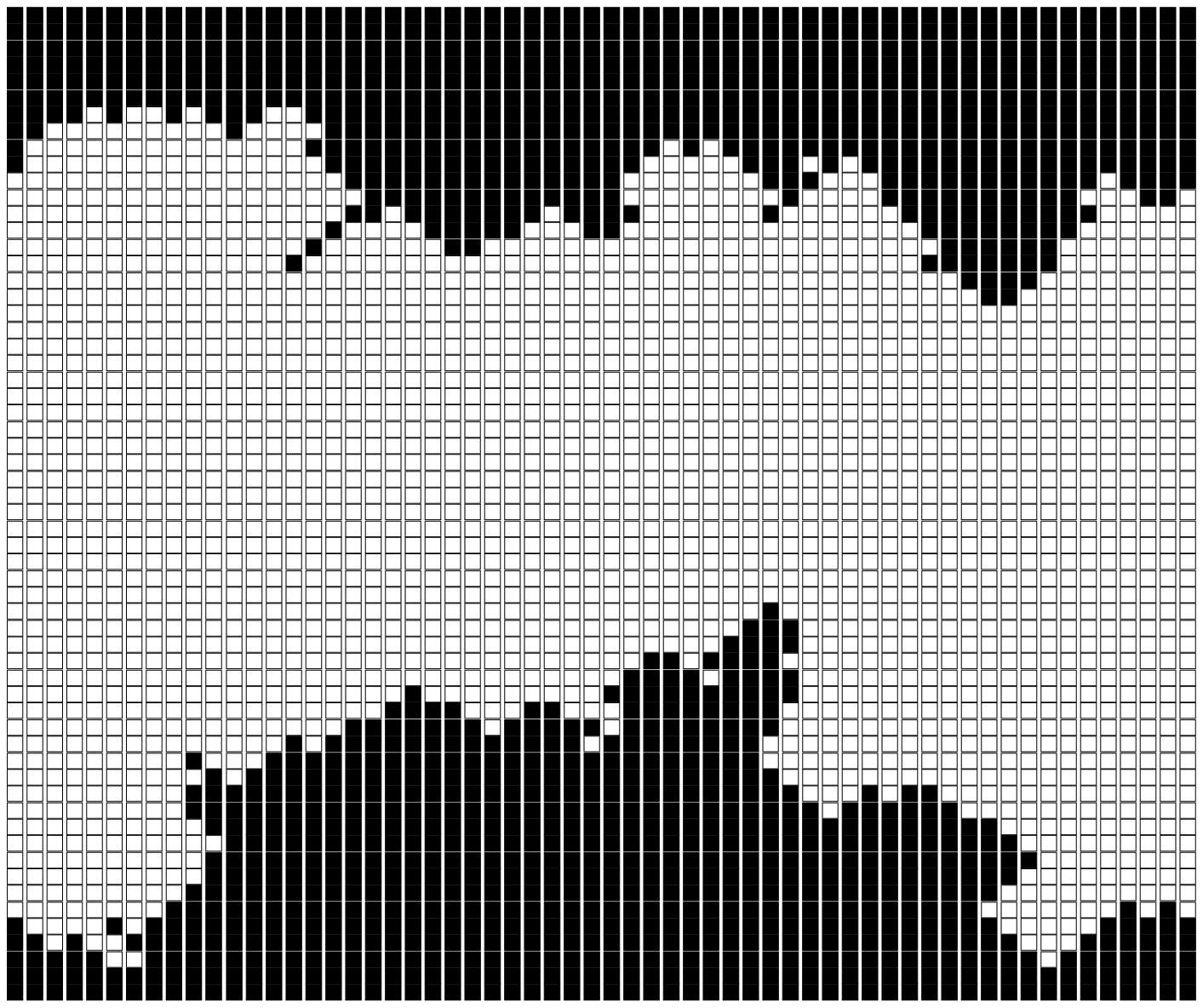}\\  \vskip -0.5ex
  \vspace*{0.2cm}
\caption{Evolution of a $60\times 60$ 9-state PR model with
  group size $G=25$ at times $t=0.1,1,4$, and 20}
\label{pics3}
\end{figure}

We now study the plurality rule (PR) model when the spatial dimension is
finite.  While the MR and PR models behave similarly in the mean-field limit,
they evolve quite differently when the spatial dimension is finite, as seen
by comparing single realizations of the system that evolve according to PR
(Fig.~\ref{pics3}) and according to MR (Fig.~\ref{pics2}), with $s=9,G=25$
for both examples.

By construction, all configurations in the PR model are active, since a
plurality exists in any group, independent of the number of states and the
group size.  Thus the PR model always evolves by diffusive domain coarsening.
When only two states remain, the ensuing evolution is exactly that of the
2-state MR model.  As in the case of the 2-state MR model, we therefore
expect that a non-negligible fraction of all realizations will get stuck in a
stripe-like state, an example of which is given in Fig.~\ref{pics3}.

\begin{figure}[ht] 
 \vspace*{0.cm}
 \includegraphics*[width=0.4\textwidth]{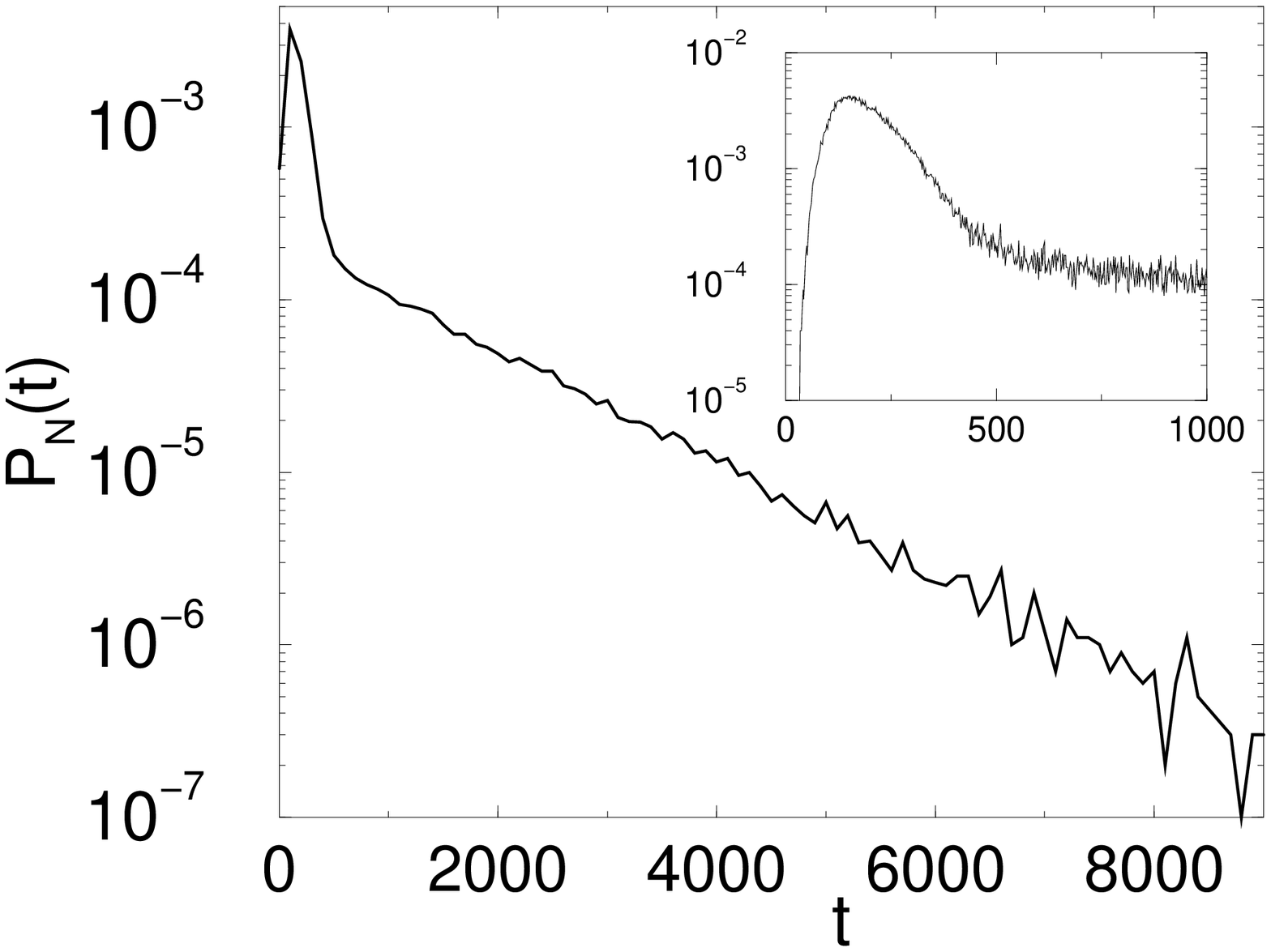}
 \includegraphics*[width=0.4\textwidth]{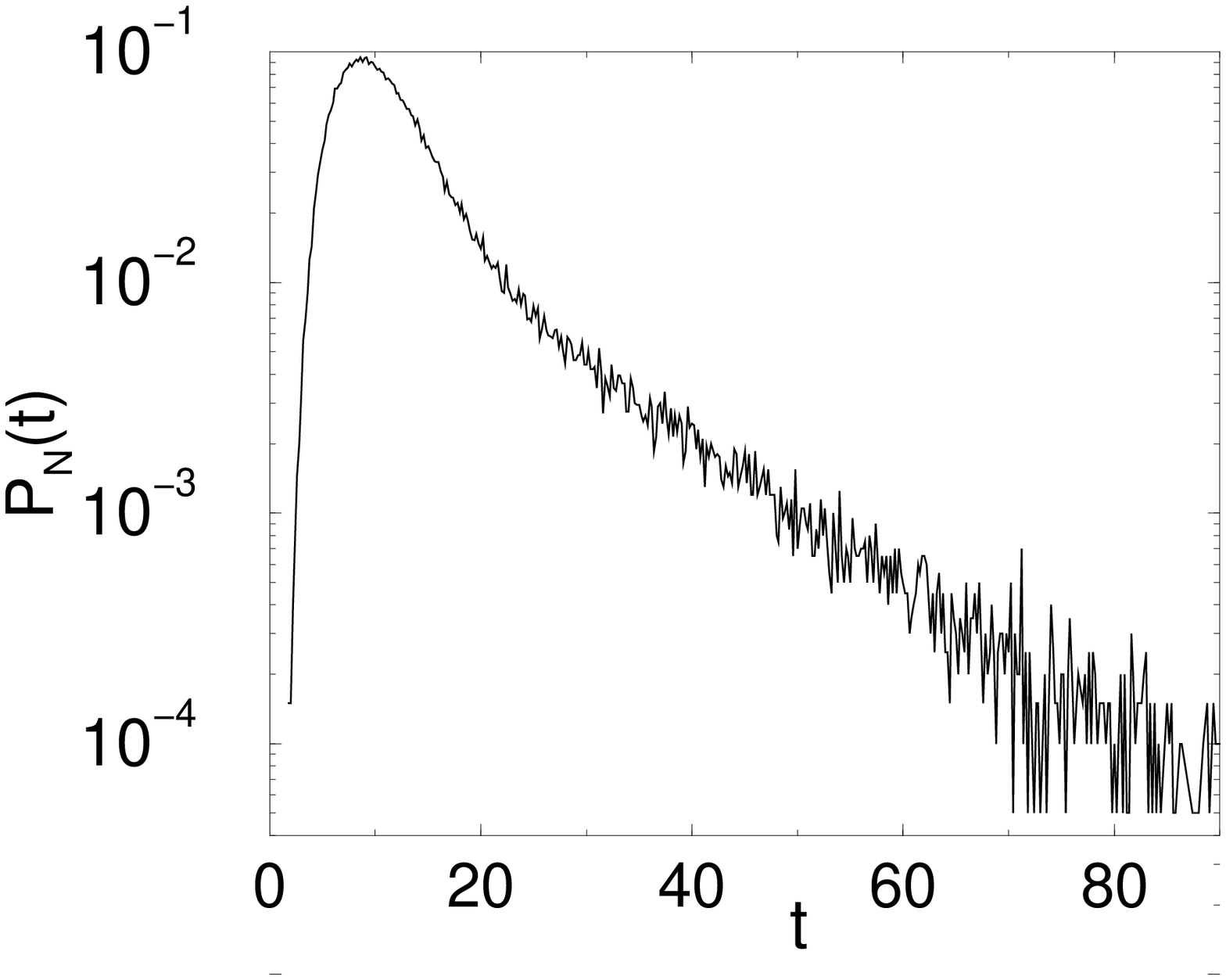}
 \caption{Probability distribution for the consensus time $P_N(t)$ versus $t$
   on a square lattice of $N=3600$ sites by plurality rule for the two cases
   of $s=3, G=3$, with data integrated over bins of width $100$ (top), and
   $s=9, G=25$, and bin size $0.2$ (bottom). The inset shows detail near the
   peak where the width of the data bins is 2.}
 \label{pdf2_pr}
\end{figure}

As a result of the correspondence with majority rule in the long-time limit,
the distribution of consensus times should again be characterized by two time
scales -- the most probable consensus time and the time associate with the
asymptotic decay of the consensus time distribution itself.  For the case of
$s=3,G=3$, the consensus time distributions in the MR and PR models are
quantitatively similar (compare the top panels of Figs.~\ref{pdf2} and
\ref{pdf2_pr}) and the behaviors of the two basic time scales in the
distribution are also nearly the same.  The actual values of $T_{\rm mp}$ and
$\tau$ in these two models are quite close, with ${T_{\rm mp}^{\rm
    PR}}/{T_{\rm mp}^{\rm MR}}=0.92$ and ${\tau^{\rm PR}}/{\tau^{\rm
    MR}}=0.98$.  As expected, the times are smaller in the PR model because
every groups is necessarily active, in contrast to the situation in majority
rule.

As the number of states is increased, the consensus time distribution in PR
continues to be described by two distinct time scales.  An example for the
case of the $s=9, G=25$ PR model is shown in Fig.~\ref{pdf2_pr}, whose
behavior strongly contrasts that of the $s=9, G=25$ MR model (bottom panel in
Fig.~\ref{pdf2}).  A final feature worth noting is that the characteristic
time scales decrease with $G$, reflecting the fact that larger groups
necessarily lead to quicker consensus formation.

\section{Summary and Discussion}

We studied two extensions of the 2-state majority rule (MR) model for
consensus formation, namely, multi-state majority rule (MR) and plurality
rule.  In the mean-field limit, both these models reach consensus in a time
that grows logarithmically with the number of agents $N$.  Variations in the
microscopic evolution rules, such as majority or plurality rule, the number
of spin states $s$, and the group size $G$, merely lead to a quantitative
difference in the amplitude of $\ln N$ in the consensus time.

For finite spatial dimension, the MR model has two distinct regimes of
behavior that are delineated by the relative values of $s$, $G$, and $N$.
For small $s$ or small $G$, many groups have local majorities in a typical
initial state.  These groups are the nuclei of domains whose long-time
dynamics is governed by diffusive coarsening.  For this range of $s$ and $G$,
the multi-state MR model thus evolves in a manner similar to that in the
$2$-state MR model.  The distribution of consensus times has two widely
separated time scales -- the most probable consensus time and the asymptotic
decay time of the consensus time distribution.  The dependences of these two
time scales on $N$ is quite close to those in the 2-state MR model.

On the other hand, for sufficiently large $s$ and $G$, it becomes
prohibitively unlikely that a finite system will contain even a initial
single group with a local majority.  Thus essentially all realizations of the
system are frozen.  As the boundary between these two regimes is approached,
a typical realization will contain either zero or one initial group with a
local majority.  In the latter case, this domain quickly imposes its state on
the entire system.  This phenomenon has a number of unfortunate historical
examples, such as Germany in 1933, Italy in 1922, and Russia in 1917, where a
well-organized, extremist, and initially minority party ultimately imposed
its will upon a deadlocked political system.

We also investigated the dynamics of plurality rule (PR), where a group
adopts the state of its plurality members in an update event.  In the
mean-field limit, PR and MR were observed to have qualitatively similar
behavior.  For finite spatial dimension, however, all groups in PR are
active.  Thus the evolution of the system for any $s$ and $G$ is
qualitatively similar to that of MR in which the number of states and the
group size is small.  As one should expect, a voting system based on
plurality rule facilitates the achievement of ultimate consensus.


\medskip

{\acknowledgments We are grateful for financial support from DOE grant
  W-7405-ENG-36 (at LANL) and NSF grant DMR0227670 (at BU).}


\begin{thebibliography}{99}
  
\bibitem{voter} T.~M.~Liggett, {\it Interacting Particle Systems}
   (Springer-Verlag, New York, 1985).
 
\bibitem{pk} 
   P. L. Krapivsky, Phys.\ Rev.\ A {\bf 45}, 1067 (1992).

\bibitem{glauber} 
   R.~J.~Glauber, J.\ Math.\ Phys.\ {\bf 4}, 294 (1963).

\bibitem{SKR} V. Spirin, P. L. Krapivsky, and S. Redner, Phys.\ Rev.\ E {\bf
     63}, 036118 (2001); Phys.\ Rev.\ E {\bf 65}, 016119 (2002).

\bibitem{galam} S. Galam, Physica (Amsterdam) {\bf 274}, 132 (1999); Eur.\ 
   Phys.\ J. B {\bf 24}, 403 (2002); {\it cond-mat}/0211571.

\bibitem{MM}
   P. L. Krapivsky and S. Redner, Phys.\ Rev.\ Lett.\ {\bf 90},  238701 (2003).
   
\bibitem{MR2D}
   P. Chen and S. Redner, Phys.\ Rev.\ E.\  {\bf 71}, 036101 (2005).
   
\bibitem{szn} K.~Sznajd-Weron and J.~Sznajd, Int.\ J. Mod.\ Phys.\ C {\bf
     11}, 1157 (2000); D.~Stauffer, J. Artif.\ Soc.\ Soc.\ Simul.\ {\bf 5},
   no.\,1 (2002).

\bibitem{GGP} A related 3-state model where tie-breaking plays a major role
   in the dynamics was recently studied by S. Gekle, S. Galam, and L. Peliti,
   {\it cond-mat}/0504254.


   



\end{thebibliography}
\end{document}